\documentclass[showpacs,aps,showkeys,graphicx,twocolumn,]{revtex4}
\usepackage{amsmath}
\usepackage{mathrsfs}
\usepackage{amsfonts}
\usepackage{amssymb}
\usepackage{graphicx}
\usepackage{eufrak}
\usepackage{multirow}
\usepackage{longtable}
\usepackage{float}
\usepackage{bm}
\usepackage{soul,color,xcolor}

\begin{document}

\title{Deterministic and complete hyperentangled Bell states analysis assisted by frequency and time interval degrees of freedom}

\author{
Xin-Jie Zhou$^{1}$, Wen-Qiang Liu$^{1,2}$, Hai-Rui Wei$^{1}$\footnote{Corresponding author: hrwei@ustb.edu.cn}, Yan-Bei Zheng$^{1}$, and Fang-Fang Du$^{3}$
}

\address{
$^1$School of Mathematics and Physics, University of Science and Technology Beijing, Beijing 100083, China\\
$^2$Center for Quantum Technology Research and Key Laboratory of Advanced Optoelectronic Quantum Architecture and Measurements (MOE), School of Physics, Beijing Institute of Technology, Beijing 100081, China\\
$^3$ Science and Technology on Electronic Test and Measurement Laboratory, North University of China, Taiyuan 030051, China\\
}

\date{\today }

\begin{abstract}
Hyperentangled Bell states analysis (HBSA) is an essential building block for certain hyper-parallel quantum information processing. We propose a complete and deterministic HBSA scheme encoded in spatial and polarization degrees of freedom (DOFs) of two-photon system assisted by a fixed frequency-based entanglement and a time interval DOF. The parity information the spatial-based and polarization-based hyper-entanglement can be distinguished by the distinct time intervals of the photon pairs, and the phase information can be distinguished by the detection signature. Compared with previous schemes, the number of the auxiliary entanglements is reduced from two to one by introducing time interval DOF. Moreover, the additional frequency and time interval DOFs suffer less from the collective channel noise.
\end{abstract}



\maketitle

\section{Introduction}\label{sec1}

Quantum entanglement \cite{book} has been considered as an essential asset to quantum communication and computation tasks. Bell states analysis (BSA), which is defined as the projection of two qubits onto maximally entangled states, is a crucial element in many important quantum information processing tasks, including measurement-based quantum computation \cite{one-way2,one-way1}, quantum teleportation \cite{teleportation2,teleportation3}, entanglement swapping \cite{swapping3,swapping2}, quantum dense coding \cite{QDC1,QDC2,QDC3}, quantum key distribution \cite{QKD6,QKD3,QKD5,QKD4,QKD-1,QKD-2,QKD-3}, entanglement concentration \cite{EC1,EC2,EC3}, and quantum secure direct communication \cite{QEDC1,QEDC2,QEDC3,QEDC4,QEDC5,QEDC6,QEDC7}. However, it is fail to unambiguously discriminate all of the four single degree of freedom (DOF) shared Bell states with only linear optical elements \cite{impossible,impossible1}, which is called a complete linear optical BSA for photonic system. Fortunately, complete and deterministic BSA can be accomplished by assorting to nonlinear interactions \cite{NMR,atom,QED}, an enlarged Hilbert space \cite{superdense}, or ancillary entangled states in additional DOFs for photonic system \cite{assisted-DOF}. The nonlinear optical tricks and the ancillary photons tricks are challenged by inefficiency and impracticality. Giant Kerr nonlinearity is a major challenge in experimental implementations.



Hyperentanglement \cite{hyperen1,hyperen2}, defined as the simultaneously entangled in multiple DOFs of a quantum system, is a fascinating resource with its high capacity, low loss rate, and some important applications (such as complete Bell-state discrimination with linear optical elements). Photons have been recognized as the excellent candidates for carrying hyperentanglement as they have a large variety of exploitable DOFs, such as polarization, frequency (wavelength), orbital angular momentum, time-bin, spatial \cite{photon-DOF2,photon-DOF1}, and the photonic qubits are much less effected by decoherence.
Among all these DOFs, the polarization DOF is the most popular candidate, sensitive to channel noise, and it can be manipulated with great precision by linear optical devices. The spatial DOF is robust against the bit-flip channel noise,  susceptible to phase error noise, and such error can be precluded by adjusting the length of spatial modes. The frequency and time interval DOFs are far more stable than polarization DOF, and frequency DOF can efficiently transfer quantum information at telecommunication wavelengths.
Nowadays, lots of researches have been devoted to the generation, manipulation, and application of hyperentanglement, and especially to hyperentangled Bell states analysis (HBSA)\cite{nonlinear8,application}.

In high-capacity quantum communication, there are $4^n$ two-photon orthogonal hyperentangled Bell (hyper-Bell) states in $n$ qubitlike DOFs. It has been shown that 16 hyper-Bell states in two DOFs can be separated into 7 groups only using linear optical elements \cite{seven1,assisted-DOF1}, and the upper bound for the number of discriminate groups in $4^n$ hyper-Bell states is $2^{n+1}-1$ \cite{assisted-DOF1}. In 2017, Li \emph{et al.} \cite{number-group1} showed that $4^n$ hyper-Bell states can be separated into $2^{n+k+1}-2^{2k}$ distinct groups via linear optics with help of $k$ ($k\leq n$) additional entangled states in ancillary DOFs. In 2019, Gao \emph{et al.} \cite{Gao-complete} proved that $2^{n+k+1}$ hyper-Bell states in $n$ ($n\geq2$) DOFs can be distinguished via linear optics with help of a time delay and $k$ ($k\leq n$) auxiliary entangled states in additional DOFs. In 2020, Gao \emph{et al.} \cite{Gao-incomplete} further enhanced to 14 distinct groups only assisted by time-bin DOF. It is note that all of above linear schemes are incomplete.
In 2010, Sheng \emph{et al.} \cite{Kerr} first proposed a complete polarization-spatial-based HBSA  with the help of cross-Kerr nonlinearity. Later some improved works were proposed \cite{nonlinear6,nonlinear7,Zhang-kerr}.
Nowadays, atoms, quantum dots, nitrogen-vacancy centers, and superconductors have been introduced to  complete deterministic HBSA \cite{nonlinear3,nonlinear4,nonlinear5,superconductor,nonlinear11,nonlinear12}. Realizing strong natural Kerr nonlinearities are challenge in experiment with current technology. The fast manipulation and measurement of neutral atoms are difficult
in experiment. $mK$ temperature and tens of $\mu$s coherence time are necessary for superconducting qubits.
In 2019, Wang \emph{et al.} \cite{Wang-complete}  proposed a program to a complete scheme in polarization and the first momentum DOFs with help of two fixed Bell states in frequency and the second momentum DOFs. Later in 2020, Zeng and Zhu \cite{Zeng-complete} completed a HBSA scheme in spatial and polarization DOFs assisted by time-bin and frequency entanglements.


In this paper, we construct a scheme for arbitrarily complete HBSA  by using frequency beam splitter, frequency shifter, and some linear optical elements. The hyper-Bell states are encoded in both the spatial and polarization DOFs of two photons, and a time intervals DOF and a fixed frequency Bell state are introduced to complete the scheme.
The parity information of the spatial and polarization DOFs (4 groups) are discriminated from each other by the time intervals of the photon pairs. The phase information of the spatial and polarization DOFs in each group can be discriminated by the detection signatures.
In contrast to the schemes in Refs. \cite{assisted-DOF,number-group1,Gao-incomplete}, the 16 hyper-Bell states are distinguished unambiguously in our scheme.
One fixed entangled state in frequency DOF is exploited to construct the present scheme which overwhelms the ones in Refs. \cite{Wang-complete,Zeng-complete}.
 In addition, frequency DOF is much more insensitive to the channel noise in an optical fiber than spatial (momentum) and polarization DOFs \cite{Wang-complete,Zeng-complete,Gao-complete}.

\section{Complete hyper-Bell states analysis in polarization and spatial DOFs of two photons} \label{sec2}

\begin{figure*}  [htbp]
\begin{center}
\includegraphics[width=15.0 cm,angle=0]{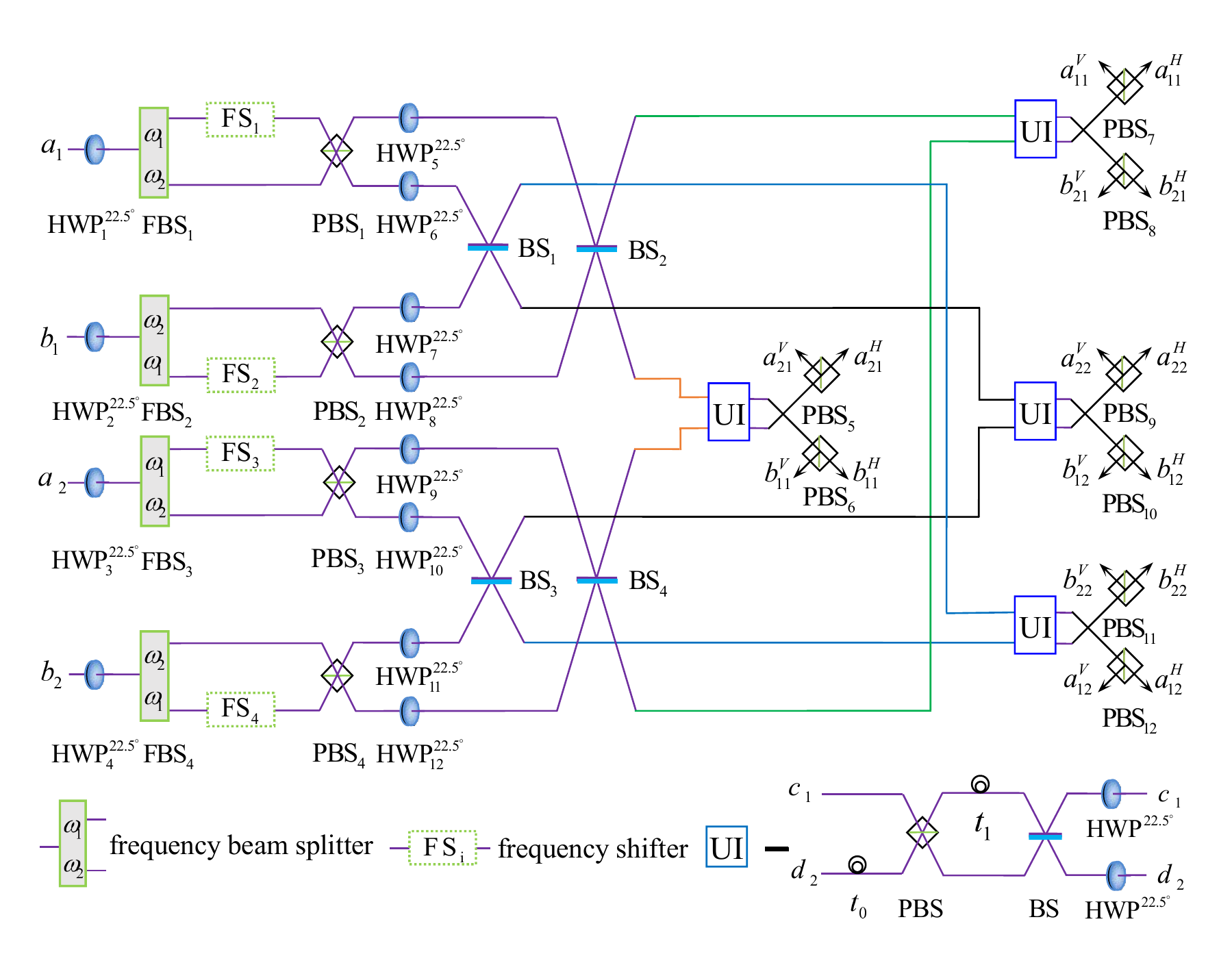}
\caption{(Color online) Schematic diagram of the complete hyper-Bell state analysis.
The frequency beam splitter (FBS) leads the photon with $\omega_1$ and $\omega_2$ into the spatial modes $x_1$ and $x_2$, respectively.
The frequency shifter (FS) completes the bit-flip operation on frequency DOF $X_f=|\omega_2\rangle\langle \omega_1| + |\omega_1\rangle\langle\omega_2|$.
UI is a unbalanced interferometer \cite{Gao-incomplete}. The circles ``circle'' on the path $c_1$ and $d_2$ introduce time intervals $t_0$ and $t_1$, respectively.
HWP$^{22.5^\circ}$ represents a half-wave plate oriented at $22.5^\circ$ of the horizontal direction, which completes the transformations $H_p=\frac{1}{\sqrt{2}}[(|H\rangle+|V\rangle)\langle H|+ (|H\rangle-|V\rangle)\langle V|]$ on the incident photon.
BS is a 50:50 beam splitter.
PBS is a polarization beam splitter, which transmits the horizontally polarized component $|H\rangle$ and reflects the vertically polarized component $|V\rangle$ of photons.
\label{Fig1}}
\end{center}
\end{figure*}


The two-photon hyper-Bell states in spatial and polarization DOFs can be written as
\begin{eqnarray}              \label{eq1}
\begin{split}
|\Lambda\rangle_{AB}=|\Theta_s\rangle_{AB}\otimes |\Gamma_p\rangle_{AB}.
\end{split}
\end{eqnarray}
We use $A$ ($B$) to denote the photon $A$ ($B$), and use the subscript $s$ ($p$) to denote the spatial (polarization) DOF.
$|\Theta _s\rangle_{AB}$ represents one of the four Bell states in spatial DOF,
\begin{eqnarray}              \label{eq2}
\begin{split}
|\phi_s^\pm\rangle_{AB} = \frac{1}{\sqrt2}(|a_1 b_1\rangle \pm |a_2 b_2\rangle),\\
|\psi_s^\pm\rangle_{AB} = \frac{1}{\sqrt2}(|a_1 b_2\rangle \pm |a_2 b_1\rangle).
\end{split}
\end{eqnarray}
$|\Gamma _p\rangle_{AB}$ represents one of the four Bell states in polarization DOF,
\begin{eqnarray}              \label{eq3}
\begin{split}
|\phi_{p}^\pm \rangle_{AB} = \frac{1}{\sqrt2}(|HH\rangle  \pm |VV\rangle),\\
| \psi_{p}^\pm \rangle_{AB} = \frac{1}{\sqrt2}(|HV\rangle  \pm |VH\rangle).
\end{split}
\end{eqnarray}
where $a_1$ ($b_1$) and $a_2$ ($b_2$) represent the two spatial modes of photon $A$ ($B$). $H$ and $V$ represent the horizontal and vertical polarization states of a single photon, respectively.
In order to completely distinguish the 16 hyper-Bell states described by Eq. (\ref{eq1}), a fixed auxiliary Bell state $|\Omega_f\rangle_{AB}$ encoded in frequency DOF is employed. $|\Omega_f\rangle_{AB}$ is given by
\begin{eqnarray}          \label{eq4}
\begin{split}
|\psi_f^+\rangle_{AB} =\frac{1}{\sqrt2}(|\omega_1\omega_2\rangle + |\omega_2 \omega_1\rangle).
\end{split}
\end{eqnarray}
Here $\omega_1$ and $\omega_2$ are the two frequencies of the incident photons.
The states of the whole system can be depicted as
\begin{eqnarray}              \label{eq5}
\begin{split}
|\Upsilon\rangle_{AB} = |\Theta_s\rangle_{AB} \otimes |\Gamma_p\rangle_{AB}\otimes|\psi_f^+\rangle_{AB}.
\end{split}
\end{eqnarray}

A scheme we designed for complete HBSA in spatial and polarization DOFs assisted by time intervals DOF and an auxiliary entanglement in frequency DOF is shown in Fig. \ref{Fig1}.
Now let us take
$|\phi_s^+\rangle_{AB} \otimes |\phi_p^+\rangle_{AB} \otimes |\psi_f^+\rangle_{AB}$,
$|\psi_s^+\rangle_{AB} \otimes |\psi_p^+\rangle_{AB} \otimes |\psi_f^+\rangle_{AB}$,
$|\phi_s^+\rangle_{AB} \otimes |\psi_p^+\rangle_{AB} \otimes |\psi_f^+\rangle_{AB}$, and $|\psi_s^+\rangle_{AB} \otimes |\phi_p^-\rangle_{AB} \otimes |\psi_f^+\rangle_{AB}$
 as examples to illustrate the principle of our scheme in detail, step by step. Here above four hyperentangled states can be written as
\begin{eqnarray}              \label{eq6}
\begin{split}
|\Upsilon_0^1\rangle=&|\phi_s^+\rangle_{AB} \otimes |\phi_p^+\rangle_{AB}\otimes |\psi_f^+\rangle_{AB}\\
                    =&\frac{1}{2\sqrt2}(|a_1 b_1\rangle + |a_2 b_2\rangle)
                               \otimes (|HH\rangle + |VV\rangle)  \\
                              &\otimes (|\omega_1\omega_2\rangle+|\omega_2\omega_{1}\rangle),
\end{split}
\end{eqnarray}
\begin{eqnarray}              \label{eq7}
\begin{split}
|\Upsilon_0^2\rangle =&|\psi_s^+\rangle_{AB} \otimes |\psi_p^-\rangle_{AB} \otimes |\psi_f^+\rangle_{AB}\\
               =&\frac{1}{2\sqrt2}(|a_1 b_2\rangle + |a_2 b_1\rangle)
                 \otimes (|HV\rangle - |VH\rangle)\\
                &\otimes (|\omega _1\omega_2\rangle + |\omega_2\omega_{1}\rangle),
\end{split}
\end{eqnarray}
\begin{eqnarray}              \label{eq8}
\begin{split}
|\Upsilon_0^3\rangle =&|\phi_s^-\rangle_{AB} \otimes |\psi_p^+\rangle_{AB} \otimes |\psi_f^+\rangle_{AB}\\
                =&\frac{1}{2\sqrt2}(|a_1 b_1\rangle - |a_2 b_2\rangle)
                          \otimes (|HV\rangle + |VH\rangle)\\
                         &\otimes (|\omega _1\omega_2\rangle + |\omega_2\omega_{1}\rangle),
\end{split}
\end{eqnarray}
\begin{eqnarray}              \label{eq9}
\begin{split}
|\Upsilon_0^4\rangle = & |\psi_s^-\rangle_{AB} \otimes |\phi_p^-\rangle_{AB} \otimes |\psi_f^+\rangle_{AB}\\
               =&\frac{1}{2\sqrt2}(|a_1 b_2\rangle - |a_2 b_1\rangle)
                          \otimes (|HH\rangle - |VV\rangle)\\
                         &\otimes (|\omega _1\omega_2\rangle + |\omega_2\omega_{1}\rangle).
\end{split}
\end{eqnarray}

First, two photons $A$ and $B$ are injected, followed by half-wave plates oriented at $22.5^\circ$ of the horizontal direction (HWP$_{1,2,3,4}^{22.5^\circ}$), frequency beam splitters (FBS$_{1,2,3,4}$), and frequency shifters (FS$_{1,2,3,4}$) in succession. Here HWP$_{1,2,3,4}^{22.5^\circ}$ accomplish the following transformations (Hadamard transformations on polarization DOF),
\begin{eqnarray}              \label{eq10}
\begin{split}
|H\rangle \xrightarrow{\text{HWP}^{22.5^\circ}_{1,2,3,4}} \frac{1}{\sqrt{2}}(|H\rangle+|V\rangle),\\
|V\rangle \xrightarrow{\text{HWP}^{22.5^\circ}_{1,2,3,4}} \frac{1}{\sqrt{2}}(|H\rangle-|V\rangle).
\end{split}
\end{eqnarray}
FS$_{1,2,3,4}$ implement the bit-flip operation on the frequency DOF, i.e.,
\begin{eqnarray}              \label{eq11}
\begin{split}
|\omega_1\rangle \xrightarrow{\text{FS}_{1,2,3,4}} |\omega_2\rangle, \qquad
|\omega_2\rangle \xrightarrow{\text{FS}_{1,2,3,4}} |\omega_1\rangle.
\end{split}
\end{eqnarray}
FBS$_{1,2,3,4}$ guide the photon to the different paths according to the frequency, i.e., the wavepackets with $\omega_1$ and $\omega_2$ are leaded to the spatial mode $x_1$ and $x_2$, respectively.
Therefore, after the operations $\text{HWP}_1^{{22.5^\circ}} \rightarrow \text{FBS}_1\rightarrow \text{FS}_1$ ($\text{HWP}_2^{{22.5^\circ}} \rightarrow \text{FBS}_2\rightarrow \text{FS}_2$, $\text{HWP}_3^{{22.5^\circ}} \rightarrow \text{FBS}_3\rightarrow \text{FS}_3$, and $\text{HWP}_4^{{22.5^\circ}} \rightarrow \text{FBS}_4\rightarrow \text{FS}_4$), the states described by Eqs. (\ref{eq6}-\ref{eq9}) become
\begin{eqnarray}              \label{eq11}
\begin{split}
|\Upsilon_1^1\rangle =&\frac{1}{2\sqrt2}(|a_1 b_1\rangle + |a_2 b_2\rangle) \otimes (|HH\rangle + |VV\rangle)\\
 &\otimes (|x_1x_2\rangle + |x_2x_{1}\rangle)\otimes|\omega_2\omega_{2}\rangle,
\end{split}
\end{eqnarray}
\begin{eqnarray}              \label{eq12}
\begin{split}
|\Upsilon_1^2\rangle =&\frac{1}{2\sqrt2}(|a_1 b_2\rangle + |a_2 b_1\rangle) \otimes (-|HV\rangle + |VH\rangle) \\
&\otimes (|x_1x_2\rangle + |x_2x_{1}\rangle)\otimes|\omega_2\omega_{2}\rangle,
\end{split}
\end{eqnarray}
\begin{eqnarray}              \label{eq13}
\begin{split}
|\Upsilon_1^3\rangle =&\frac{1}{2\sqrt2}(|a_1 b_1\rangle - |a_2 b_2\rangle) \otimes (|HH\rangle - |VV\rangle) \\
&\otimes (|x_1x_2\rangle + |x_2x_{1}\rangle)\otimes|\omega_2\omega_{2}\rangle,
\end{split}
\end{eqnarray}
\begin{eqnarray}              \label{eq14}
\begin{split}
|\Upsilon_1^4\rangle =&\frac{1}{2\sqrt2}(|a_1 b_2\rangle - |a_2 b_1\rangle) \otimes (|HV\rangle + |VH\rangle)\\
 &\otimes (|x_1x_2\rangle + |x_2x_{1}\rangle)\otimes|\omega_2\omega_{2}\rangle.
\end{split}
\end{eqnarray}

Second,  after the photons pass through polarization beam splitters (PBS$_{1,2,3,4}$), and HWP$^{22.5^\circ}_{5,6,7,8,9,10,11,12}$  in succession, the states described by Eqs. (\ref{eq11}-\ref{eq14}) then become
\begin{eqnarray}              \label{eq15}
\begin{split}
|\Upsilon_2^1\rangle=&\frac{1}{2\sqrt2}(|a_1 b_1\rangle + |a_2 b_2\rangle) \otimes (|HH\rangle + |VV\rangle) \\
&\otimes (|{x_1}{x_2}\rangle  + |{x_2}{x_1}\rangle )
\otimes|{\omega _2}{\omega _2}\rangle,
\end{split}
\end{eqnarray}
\begin{eqnarray}              \label{eq16}
\begin{split}
|\Upsilon_2^2\rangle=&\frac{1}{2\sqrt2}(|a_1 b_2\rangle + |a_2 b_1\rangle) \otimes (-|HV\rangle + |VH\rangle) \\
&\otimes (|x_1x_1\rangle + |x_2x_2\rangle)\otimes|\omega_2\omega_{2}\rangle,
\end{split}
\end{eqnarray}
\begin{eqnarray}              \label{eq17}
\begin{split}
|\Upsilon_2^3\rangle=&\frac{1}{2\sqrt2}(|a_1 b_1\rangle - |a_2 b_2\rangle) \otimes (|HV\rangle + |VH\rangle)\\
 &\otimes (|x_1x_2\rangle + |x_2x_{1}\rangle)\otimes|\omega_2\omega_{2}\rangle,
\end{split}
\end{eqnarray}
\begin{eqnarray}              \label{eq18}
\begin{split}
|\Upsilon_2^4\rangle=&\frac{1}{2\sqrt2}(|a_1 b_2\rangle - |a_2 b_1\rangle) \otimes (|HH\rangle - |VV\rangle)\\
&\otimes (|x_1x_1\rangle + |x_2x_2\rangle)\otimes|\omega_2\omega_{2}\rangle.
\end{split}
\end{eqnarray}

Third, the wave-packets mixed at 50:50 nonpolarization  beam splitters (BS$_{1,2,3,4}$), and then the states will be evolved to
\begin{eqnarray}              \label{eq19}
\begin{split}
|\Upsilon_3^1\rangle=&\frac{1}{4\sqrt2}
  (|{a_1}{a_1}\rangle  - |{a_1}{b_1}\rangle
  + |{b_1}{a_1}\rangle  - |{b_1}{b_1}\rangle  \\
  &+ |{a_2}{a_2}\rangle  - |{a_2}{b_2}\rangle
  + |{b_2}{a_2}\rangle  - |{b_2}{b_2}\rangle) \\
  & \otimes (|HH\rangle  + |VV\rangle) \otimes (|{x_1}{x_2}\rangle  + |{x_2}{x_1}\rangle)\\&
  \otimes|{\omega _2}{\omega _2}\rangle,
\end{split}
\end{eqnarray}
\begin{eqnarray}              \label{eq20}
\begin{split}
|\Upsilon_3^2\rangle=&\frac{1}{4\sqrt2}
  (|{a_1}{a_2}\rangle  - |a_1b_2\rangle
  + |b_1a_2\rangle  - |b_1b_2\rangle  \\
  &+ |a_2a_1\rangle  - |a_2b_1\rangle
  + |b_2a_1\rangle  - |b_2b_1\rangle) \\
  & \otimes (|HV\rangle  - |VH\rangle)\otimes (|x_1x_1\rangle  + |x_2x_2\rangle)\\&
 \otimes|{\omega _2}{\omega _2}\rangle,
\end{split}
\end{eqnarray}
\begin{eqnarray}              \label{eq21}
\begin{split}
|\Upsilon_3^3\rangle=&\frac{1}{4\sqrt2}
  (|{a_1}{a_1}\rangle  - |{a_1}{b_1}\rangle
  + |{b_1}{a_1}\rangle   - |{b_1}{b_1}\rangle \\
 &  - |{a_2}{a_2}\rangle  + |{a_2}{b_2}\rangle
   - |{b_2}{a_2}\rangle  + |{b_2}{b_2}\rangle ) \\
   &\otimes (|HV\rangle  + |VH\rangle)
   \otimes (|{x_1}{x_2}\rangle  + |{x_2}{x_1}\rangle)\\&
   \otimes|{\omega _2}{\omega _2}\rangle,
\end{split}
\end{eqnarray}
\begin{eqnarray}              \label{eq22}
\begin{split}
|\Upsilon_3^4\rangle=&\frac{1}{4\sqrt2}
  (|a_1a_2\rangle  - |{a_1}{b_2}\rangle
  + |{b_1}{a_2}\rangle  - |{b_1}{b_2}\rangle\\
 & - |{a_2}{a_1}\rangle  + |{a_2}{b_1}\rangle
   - |{b_2}{a_1}\rangle  + |{b_2}{b_1}\rangle )\\
  & \otimes (|HH\rangle  - |VV\rangle)\otimes (|{x_1}{x_1}\rangle  + |{x_2}{x_2}\rangle)\\&
  \otimes|{\omega _2}{\omega _2}\rangle.
\end{split}
\end{eqnarray}
Here the operations of 50:50 BS are given by
\begin{eqnarray}              \label{eq24}
\begin{split}
a_{1(2)}\xrightarrow{\text{BS}}\frac{1}{\sqrt{2}}(a_{1(2)}+ b_{1(2)}),\\
b_{1(2)}\xrightarrow{\text{BS}}\frac{1}{\sqrt{2}}(a_{1(2)}- b_{1(2)}).
\end{split}
\end{eqnarray}
Fourth, the photons are sent into the unbalanced interferometers (UIs) block \cite{Gao-incomplete} (see Fig. \ref{Fig1}), which is used to complete the transformations
\begin{eqnarray}              \label{eq25}
\begin{split}
|H\rangle|m_1\rangle\rightarrow&I(|H\rangle|m_1\rangle  + |V\rangle |m_1\rangle - |H\rangle|n_2\rangle \\ & - |V\rangle |n_2\rangle ),\\
|V\rangle|n_2\rangle\;\rightarrow&D(t_0)(|H\rangle|m_1\rangle  - |V\rangle|m_1\rangle- |H\rangle |n_2\rangle  \\&+ |V\rangle|n_2\rangle ),\\
|V\rangle|m_1\rangle\rightarrow& D(t_1)(|H\rangle|m_1\rangle  + |V\rangle|m_1\rangle
- |H\rangle |n_2\rangle  \\& - |V\rangle|n_2\rangle),\\
|H\rangle|n_2\rangle\rightarrow&D(t_0+t_1)(|H\rangle|m_1\rangle
+ |V\rangle|m_1\rangle + |H\rangle |n_2\rangle \\& + |V\rangle |n_2\rangle).\\
\end{split}
\end{eqnarray}
Here $m$ and $n$ represent the upper and down arms of the UI block, respectively.
$I=D(0)$, $D(t_0)$, $D(t_1)$, and $D(t_0 + t_1)$ are linear time delay operators (time intervals) performed on the incident photons \cite{Gao-incomplete,superdense,linear time}.
In the scheme, $\omega t_0 = 2n\pi$ and $\omega t_1 = 2m\pi$ ($n$ and $m$ are integers, and $\omega$ is the frequency of input photon) should be taken to make the photons interfere with each other under the constructive-interfere condition \cite{Gao-incomplete,superdense}. It is noted that $t_1\pm t_0$ must be indistinguishable in the construction, that is, if the time resolution of the detectors is 4 $ns$, the time intervals $t_0$ and $t_1$ in UI could be set to 6 $ns$ and 10 $ns$ in experiment, respectively \cite{Gao-incomplete,superdense,linear time}.
Hence, after the photons interact with the UIs, the states described by Eqs. (\ref{eq19}-\ref{eq22}) evolve as
\begin{eqnarray}              \label{eq27}
\begin{split}
|\Upsilon_4^1\rangle=&\frac{1}{4}[II(|{b_1}{a_2} \rangle+ |{a_2}{b_1} \rangle- |{a_1}{b_2} \rangle- |{b_2}{a_1}\rangle)  \\
&\otimes (|HV\rangle  + |VH\rangle)]\otimes (|{x_1}{x_2}\rangle  + |{x_2}{x_1}\rangle)\\
&\otimes|{\omega _2}{\omega _2}{\rangle},
\end{split}
\end{eqnarray}
\begin{eqnarray}              \label{eq28}
\begin{split}
|\Upsilon_4^2\rangle=&\frac{1}{8}
 [ ID({T_0})(( -| {a_1}{a_2}\rangle -| {b_2}{b_1}\rangle + |{b_1}{b_2} \rangle
 \\   &+ |{a_2}{a_1}\rangle)
 \otimes (|HH\rangle  - |VV\rangle )\\
   &-ID({T_0})(|{a_1}{b_1}\rangle+ |{b_2}{a_2}\rangle -| {b_1}{a_1}\rangle - |{a_2}{b_2}\rangle)   \\
  & \otimes (|HV\rangle  - |VH\rangle) \\
   &+  D({T_0})I(|({b_1}{b_2}\rangle + |{a_2}{a_1}\rangle
   - | {a_1}{a_2}\rangle -| {b_2}{b_1})\rangle \\   &   \otimes (|HH\rangle  - |VV\rangle )\\
   &-D({T_0})I |({b_1}{a_1}\rangle + |{a_2}{b_2}\rangle -| {a_1}{b_1}\rangle - | {b_2}{a_2}\rangle)\\
    & \otimes (|HV\rangle  - |VH\rangle)] \otimes (|{x_1}{x_1}\rangle  + |{x_2}{x_2}\rangle)\\
    &\otimes|{\omega _2}{\omega _2}{\rangle},
\end{split}
\end{eqnarray}
\begin{eqnarray}              \label{eq28}
\begin{split}
|\Upsilon_4^3\rangle=&\frac{1}{8}
[ ID({T_1})(|{a_1}{a_1}\rangle - |{b_2}{b_2}\rangle +|{a_1}{b_2}\rangle -| {b_2}{a_1}\rangle)\\
&\otimes (|HH\rangle  - |VV\rangle)\\
 &- ID({T_1})(|{b_1}{b_1}\rangle +| {b_1}{a_2}\rangle -| {a_2}{b_1}\rangle + |{a_2}{a_2}\rangle) \\
  &\otimes (|HH\rangle  - |VV\rangle)\\
 &  - D({T_1})I(|{b_1}{b_1}\rangle - |{a_2}{a_2}\rangle - |{b_1}{a_2}\rangle +| {a_2}{b_1}\rangle)\\
   & \otimes (|HH\rangle  - |VV\rangle )   \\
   &+ D({T_1})I(|{a_1}{a_1}\rangle - |{b_2}{b_2}\rangle + |{b_2}{a_1}\rangle - |{a_1}{b_2}\rangle)\\
  & \otimes (|HH\rangle  - |VV\rangle )]\otimes (|{x_1}{x_2}\rangle  + |{x_2}{x_1}\rangle)\\  &
  \otimes|{\omega _2}{\omega _2}{\rangle},
\end{split}
\end{eqnarray}
\begin{eqnarray}              \label{eq30}
\begin{split}
|\Upsilon_4^4\rangle=&\frac{1}{8\sqrt2}
  [ID({t_0} + {t_1})(-|  {a_1}{a_1}\rangle  - |{a_1}{b_2}\rangle+|{b_2}{a_1}\rangle \\
  & +| {b_2}{b_2 } \rangle)
\otimes (|HH\rangle  + |HV\rangle  + |VH\rangle  + |VV\rangle)  \\
   &+ID({t_0} + {t_1})( |{b_1}{b_1}\rangle+|{b_1}{a_2}\rangle  -| {a_2}{b_1}\rangle
   \\
 & - |{a_2}{a_2}\rangle ) \otimes (|HH\rangle  + |HV\rangle  + |VH\rangle  + |VV\rangle ) \\
   &+ D(t_1)D(t_0)( |{a_1}{a_1}\rangle - |{a_1}{b_2}\rangle + | {b_2}{a_1}\rangle \\  &
   -|{b_2}{b_2}\rangle)\otimes (|HH\rangle  - |HV\rangle  - |VH\rangle  + |VV\rangle)\\
  & + D(t_1)D(t_0)(|{b_1}{b_1}\rangle - |{b_1}{a_2}\rangle +| {a_2}{b_1}\rangle\\
  &
  - | {a_2}{a_2}\rangle)   \otimes (|HH\rangle  - |HV\rangle  - |VH\rangle  + |VV\rangle)]
  \\&
  \otimes (|{x_1}{x_1}\rangle  + |{x_2}{x_2}\rangle)
  \otimes|{\omega _2}{\omega _2}{\rangle}.
\end{split}
\end{eqnarray}


\begin{table*}[htb]
	\centering
	\caption{Relations between the 16 hyperentangled  Bell states, the detection signatures, and the time intervals. $0$, $t_0$, $t_1$, and $t_1 \pm t_0$ are the time intervals of the two outing photons. }
	\begin{tabular}{ccc}
		\hline  \hline
Input       & Detection             & Time \\
states      & signatures            & intervals  \\ \hline

	$|\phi^+_s\rangle\otimes |\phi^+_p\rangle$ &
	$a_{11}^{H(V)}b_{22}^{V(H)}$, $a_{12}^{H(V)}b_{21}^{V(H)}$, $b_{11}^{H(V)}a_{22}^{V(H)}$, $b_{12}^{H(V)}a_{21}^{V(H)}$ 	 &\multirow{4}{*}{0}\\ \cline{2-2}

	$|{\phi^+_s}\rangle\otimes|{\phi^-_p}\rangle$&
   $a_{11}^{H(V)}b_{21}^{H(V)}$, $a_{12}^{H(V)}b_{22}^{H(V)}$, $b_{11}^{H(V)}a_{21}^{H(V)}$, $b_{12}^{H(V)}a_{22}^{H(V)}$\\\cline{2-2}
	
	$|{\phi^-_s}\rangle\otimes|{\phi^+_p}\rangle$&
	$a_{11}^{H(V)}a_{12}^{H(V)}$, $a_{12}^{H(V)}a_{11}^{H(V)}$, $a_{21}^{H(V)}a_{22}^{H(V)}$, $a_{22}^{H(V)}a_{21}^{H(V)}$, 
	$b_{11}^{H(V)}b_{12}^{H(V)}$, $b_{12}^{H(V)} b_{11}^{H(V)}$, $b_{21}^{H(V)} b_{22}^{H(V)}$, $b_{22}^{H(V)}b_{21}^{H(V)}$\\\cline{2-2}
	
	$|{\phi^-_s}\rangle\otimes|{\phi^-_p}\rangle$&
	$a_{11}^{H(V)}a_{11}^{V(H)}$, $a_{12}^{H(V)}a_{12}^{V(H)}$, $a_{21}^{H(V)}a_{21}^{V(H)}$, $a_{22}^{H(V)}a_{22}^{V(H)}$,
	$b_{11}^{H(V)} b_{11}^{V(H)}$, $b_{12}^{H(V)} b_{12}^{V(H)}$, $b_{21}^{H(V)} b_{21}^{V(H)}$, $ b_{22}^{H(V)} b_{22}^{V(H)}$ \\
	\hline
	
	$|{\psi^+_s}\rangle\otimes|{\psi^+_p}\rangle$&
	$a_{11}^{H(V)}b_{12}^{H(V)}$, $a_{12}^{H(V)}b_{11}^{H(V)}$, $a_{21}^{H(V)}b_{22}^{H(V)}$, $a_{22}^{H(V)}b_{21}^{H(V)}$,
	$a_{11}^{H(V)}a_{22}^{V(H)}$,  $a_{12}^{H(V)}a_{21}^{V(H)}$, $b_{11}^{H(V)}b_{22}^{V(H)}$, $b_{12}^{H(V)}b_{21}^{V(H)}$ & \multirow{6}{*}{$t_0$} \\\cline{2-2}
	
	$|{\psi^+_s}\rangle\otimes|{\psi^-_p}\rangle$&
	$a_{11}^{H(V)}b_{11}^{V(H)}$, $a_{12}^{H(V)}b_{12}^{V(H)}$, $a_{21}^{H(V)}b_{21}^{V(H)}$, $a_{22}^{H(V)}b_{22}^{V(H)}$,
	$a_{11}^{H(V)}a_{21}^{H(V)}$,  $a_{12}^{H(V)}a_{22}^{H(V)}$, $b_{11}^{H(V)}b_{21}^{H(V)}$, $ b_{12}^{H(V)}b_{22}^{H(V)}$ \\\cline{2-2}
	
	\multirow{2}{*}{$|{\psi^-_s}\rangle\otimes|{\psi^+_p}\rangle$}
	&$a_{11}^{H(V)}a_{12}^{H(V)}$, $a_{12}^{H(V)}a_{11}^{H(V)}$, $a_{21}^{H(V)}a_{22}^{H(V)}$, $a_{22}^{H(V)}a_{21}^{H(V)}$,\\&	
	$b_{11}^{H(V)}b_{12}^{H(V)}$,  $b_{12}^{H(V)}b_{11}^{H(V)}$, $b_{21}^{H(V)}b_{22}^{H(V)}$, $b_{22}^{H(V)}b_{21}^{H(V)}$,
	$a_{11}^{H(V)}b_{22}^{V(H)}$,  $a_{12}^{H(V)}b_{21}^{V(H)}$, $a_{21}^{H(V)}b_{12}^{V(H)}$, $a_{22}^{H(V)} b_{11}^{V(H)}$ \\\cline{2-2}
	
	\multirow{2}{*}{$|\psi^-_s\rangle\otimes|\psi^-_p\rangle$}
	&$a_{11}^{H(V)}a_{11}^{V(H)}$, $a_{12}^{H(V)}a_{12}^{V(H)}$,  $a_{21}^{H(V)}a_{21}^{V(H)}$, $a_{22}^{H(V)}a_{22}^{V(H)}$,\\&	
	$b_{11}^{H(V)} b_{11}^{V(H)}$, $b_{12}^{H(V)} b_{12}^{V(H)}$, $b_{21}^{H(V)}b_{21}^{V(H)}$, $b_{22}^{H(V)}b_{22}^{V(H)}$,
	$a_{11}^{H(V)}b_{21}^{H(V)}$, $a_{12}^{H(V)} b_{22}^{H(V)}$,  $a_{21}^{H(V)}b_{11}^{H(V)}$, $a_{22}^{H(V)}b_{12}^{H(V)}$ \\\cline{2-2}
	\hline
	
	\multirow{2}{*}{$|\phi^+_s\rangle\otimes|\psi^+_p\rangle$}
	&$a_{11}^{H(V)}a_{12}^{V(H)}$, $a_{12}^{H(V)}a_{11}^{V(H)}$, $a_{21}^{H(V)}a_{22}^{V(H)}$, $a_{22}^{H(V)}a_{21}^{V(H)}$,\\&	
	$b_{11}^{H(V)}b_{12}^{V(H)}$, $b_{12}^{H(V)}b_{11}^{V(H)}$, $b_{21}^{H(V)}b_{22}^{V(H)}$, $b_{22}^{H(V)}b_{21}^{V(H)}$,
	$a_{11}^{H(V)}b_{22}^{V(H)}$, $a_{12}^{H(V)}b_{21}^{V(H)}$, $a_{21}^{H(V)}b_{12}^{V(H)}$, $a_{22}^{H(V)}b_{11}^{V(H)}$ & \multirow{6}{*}{$t_1$}\\\cline{2-2}
	
	\multirow{2}{*}{$|{\phi^+_s}\rangle\otimes|{\psi^-_p}\rangle$}
	&$a_{11}^{H(V)}a_{11}^{H(V)}$, $a_{12}^{H(V)}a_{12}^{H(V)}$, $a_{21}^{H(V)}a_{21}^{H(V)}$, $a_{22}^{H(V)}a_{22}^{H(V)}$,\\&	
	$b_{11}^{H(V)}b_{11}^{H(V)}$, $b_{12}^{H(V)}b_{12}^{H(V)}$, $b_{21}^{H(V)}b_{21}^{H(V)}$, $b_{22}^{H(V)}b_{22}^{H(V)}$,
	$a_{11}^{H(V)}b_{21}^{H(V)}$, $a_{12}^{H(V)}b_{22}^{H(V)}$, $a_{21}^{H(V)}b_{11}^{H(V)}$, $a_{22}^{H(V)}b_{12}^{H(V)}$  \\\cline{2-2}
	
	\multirow{2}{*}{$|{\phi^-_s}\rangle\otimes|{\psi^+_p}\rangle$}
	&$a_{11}^{H(V)}a_{12}^{H(V)}$, $a_{12}^{H(V)}a_{11}^{H(V)}$, $a_{21}^{H(V)}a_{22}^{H(V)}$, $a_{22}^{H(V)}a_{21}^{H(V)}$,\\&	
	$b_{11}^{H(V)}b_{12}^{H(V)}$,  $b_{12}^{H(V)}b_{11}^{H(V)}$, $b_{21}^{H(V)}b_{22}^{H(V)}$, $b_{22}^{H(V)}b_{21}^{H(V)}$,
	$a_{11}^{H(V)}b_{22}^{H(V)}$,  $a_{12}^{H(V)}b_{21}^{H(V)}$, $a_{21}^{H(V)}b_{12}^{H(V)}$, $a_{22}^{H(V)}b_{11}^{H(V)}$ \\\cline{2-2}
	
	\multirow{2}{*}{$|{\phi^-_s}\rangle\otimes|{\psi^-_p}\rangle$}
	&$a_{11}^{H(V)}a_{11}^{V(H)}$, $a_{12}^{H(V)}a_{12}^{V(H)}$, $a_{21}^{H(V)}a_{21}^{V(H)}$, $a_{22}^{H(V)}a_{22}^{V(H)}$,\\&	
	$b_{11}^{H(V)}b_{11}^{V(H)}$, $b_{12}^{H(V)}b_{12}^{V(H)}$, $b_{21}^{H(V)}b_{21}^{V(H)}$, $b_{22}^{H(V)}b_{22}^{V(H)}$,
	$a_{11}^{H(V)}b_{21}^{V(H)}$, $a_{12}^{H(V)}b_{22}^{V(H)}$, $a_{21}^{H(V)}b_{11}^{V(H)}$, $a_{22}^{H(V)}b_{12}^{V(H)}$ \\\cline{2-2}
	\hline
	
	\multirow{2}{*}{$|{\psi^+_s}\rangle\otimes|{\phi^+_p}\rangle$}
	&$a_{11}^{H(V)}a_{22}^{H(V)}$, $a_{11}^{H(V)}a_{22}^{V(H)}$, $a_{12}^{H(V)}a_{21}^{H(V)}$, $a_{12}^{H(V)}a_{21}^{V(H)}$,
	$b_{11}^{H(V)}b_{22}^{H(V)}$, $b_{11}^{H(V)}b_{22}^{V(H)}$, $b_{12}^{H(V)}b_{21}^{H(V)}$, $b_{12}^{H(V)}b_{21}^{V(H)}$,\\&		
	$a_{11}^{H(V)}b_{12}^{H(V)}$, $a_{11}^{H(V)}b_{12}^{V(H)}$, $a_{12}^{H(V)}b_{11}^{H(V)}$, $a_{12}^{H(V)}b_{11}^{V(H)}$,
	$a_{21}^{H(V)}b_{22}^{H(V)}$, $a_{21}^{H(V)}b_{22}^{V(H)}$, $a_{22}^{H(V)}b_{21}^{H(V)}$, $a_{22}^{H(V)}b_{21}^{V(H)}$ & \multirow{10}{*}{$t_1\pm t_0$}\\\cline{2-2}
	
	\multirow{2}{*}{$|{\psi^+_s}\rangle\otimes|{\phi^-_p}\rangle$}	&
	$a_{11}^{H(V)}a_{21}^{H(V)}$, $a_{11}^{H(V)}a_{21}^{V(H)}$, $a_{12}^{H(V)}a_{22}^{H(V)}$, $a_{12}^{H(V)}a_{22}^{V(H)}$,
	$b_{11}^{H(V)}b_{21}^{H(V)}$, $b_{12}^{H(V)}b_{22}^{H(V)}$, $b_{11}^{H(V)}b_{21}^{V(H)}$, $b_{12}^{H(V)}b_{22}^{V(H)}$,\\&
	$a_{11}^{H(V)}b_{11}^{H(V)}$, $a_{11}^{H(V)}b_{11}^{V(H)}$, $a_{12}^{H(V)}b_{12}^{H(V)}$, $a_{12}^{H(V)}b_{12}^{V(H)}$,
	$a_{21}^{H(V)}b_{21}^{H(V)}$, $a_{21}^{H(V)}b_{21}^{V(H)}$, $a_{22}^{H(V)}b_{22}^{H(V)}$, $a_{22}^{H(V)}b_{22}^{V(H)}$	\\\cline{2-2}
	
	\multirow{3}{*}{$|{\psi^-_s}\rangle\otimes|{\phi^+_p}\rangle$}
	&$a_{11}^{H(V)}a_{12}^{H(V)}$, $a_{12}^{H(V)}a_{11}^{H(V)}$, $a_{21}^{H(V)}a_{22}^{H(V)}$, $a_{22}^{H(V)}a_{21}^{H(V)}$,
	$b_{11}^{H(V)}b_{12}^{H(V)}$, $b_{12}^{H(V)}b_{11}^{H(V)}$, $b_{21}^{H(V)}b_{22}^{H(V)}$, $b_{22}^{H(V)}b_{21}^{H(V)}$,\\&
	$a_{11}^{H(V)}b_{22}^{H(V)}$, $a_{12}^{H(V)}b_{21}^{H(V)}$, $a_{11}^{H(V)}b_{22}^{V(H)}$, $a_{12}^{H(V)}b_{21}^{V(H)}$,
	$a_{21}^{H(V)}b_{12}^{H(V)}$, $a_{22}^{H(V)}b_{11}^{H(V)}$, $a_{21}^{H(V)}b_{12}^{V(H)}$, $a_{22}^{H(V)}b_{11}^{V(H)}$,\\&
	$a_{11}^{H(V)}a_{12}^{V(H)}$, $a_{12}^{H(V)}a_{11}^{V(H)}$, $a_{21}^{H(V)}a_{22}^{V(H)}$, $a_{22}^{H(V)}a_{21}^{V(H)}$, 
	$b_{11}^{H(V)}b_{12}^{V(H)}$, $b_{12}^{H(V)}b_{11}^{V(H)}$, $b_{21}^{H(V)}b_{22}^{V(H)}$, $b_{22}^{H(V)}b_{21}^{V(H)}$ \\\cline{2-2}
	
	\multirow{3}{*}{$|{\psi^-_s}\rangle\otimes|{\phi^-_p}\rangle$}
	&$a_{11}^{H(V)}a_{11}^{H(V)}$, $a_{11}^{H(V)}a_{11}^{V(H)}$, $a_{12}^{H(V)}a_{12}^{H(V)}$, $a_{12}^{H(V)}a_{12}^{V(H)}$,
	$a_{21}^{H(V)}a_{21}^{H(V)}$, $a_{21}^{H(V)}a_{21}^{V(H)}$, $a_{22}^{H(V)}a_{22}^{H(V)}$, $a_{22}^{H(V)}a_{22}^{V(H)}$,\\&
	$b_{11}^{H(V)}b_{11}^{H(V)}$, $b_{11}^{H(V)}b_{11}^{V(H)}$, $b_{12}^{H(V)}b_{12}^{H(V)}$, $b_{12}^{H(V)}b_{12}^{V(H)}$,
	$b_{21}^{H(V)}b_{21}^{H(V)}$, $b_{21}^{H(V)}b_{21}^{V(H)}$, $b_{22}^{H(V)}b_{22}^{H(V)}$, $b_{22}^{H(V)}b_{22}^{V(H)}$,\\&
	$a_{11}^{H(V)}b_{21}^{H(V)}$, $a_{12}^{H(V)}b_{22}^{H(V)}$, $a_{11}^{H(V)}b_{21}^{V(H)}$, $a_{12}^{H(V)}b_{22}^{V(H)}$, 
	$a_{21}^{H(V)}b_{11}^{H(V)}$, $a_{21}^{H(V)}b_{11}^{V(H)}$, $a_{22}^{H(V)}b_{12}^{H(V)}$, $a_{22}^{H(V)}b_{12}^{V(H)}$ \\
	\hline  \hline
\end{tabular}\label{table1}
\end{table*}

Finally, the two output photons are detected in the $\{|H\rangle, |V\rangle\}$ basis by using PBSs and single-photon detectors. Here polarization beam splitter, PBS, transmits the $H$-polarized component and reflects the $V$-polarized component, respectively. As shown in Tab. \ref{table1}, the 16 distinct detection signatures correspond to the 16 hyper-Bell states, respectively. For example, the state $|{\psi^+_s}\rangle\otimes|{\phi^-_p}\rangle$ will trigger one of the detection signatures for two detectors
$\{
a_{11}^{H(V)}b_{11}^{H(V)}$, $a_{12}^{H(V)}b_{12}^{H(V)}$,
$a_{11}^{H(V)}a_{21}^{H(V)}$, $a_{12}^{H(V)}a_{22}^{H(V)}$,
$b_{21}^{H(V)}b_{11}^{H(V)}$, $b_{22}^{H(V)}b_{12}^{H(V)}$,
$b_{21}^{H(V)}a_{21}^{H(V)}$, $b_{22}^{H(V)}a_{22}^{H(V)}$,
$a_{11}^{H(V)}b_{11}^{V(H)}$, $a_{12}^{H(V)}b_{12}^{V(H)}$,
$a_{11}^{H(V)}a_{21}^{V(H)}$, $a_{12}^{H(V)}a_{22}^{V(H)}$,
$b_{21}^{H(V)}b_{11}^{V(H)}$, $b_{22}^{H(V)}b_{12}^{V(H)}$,
$b_{21}^{H(V)}a_{21}^{V(H)}$, $b_{22}^{H(V)}a_{22}^{V(H)}\}$
with time intervals ($t_1\pm t_0$). The discriminate time intervals of the photon-pair divide the 16 hyper-Bell states into 4 groups, and the detection signatures in each group are distincted from each other.

Based on above discussion, one can see that the scheme shown in Fig. \ref{Fig1} completely distinguishes 16 hyper-Bell states encoded in spatial and polarization DOFs of two-photon. The distinction of the parity information of the spatial and polarization DOFs (be divided into 4 groups) resorts to the time intervals of the photon pairs. The distinction of the phase information of spatial and polarization DOFs resorts to the diacritical detection signatures.


\section{Conclusion} \label{sec4}

In this paper, we theoretically presented a scheme to completely distinguish 16 hyper-Bell states encoded in the spatial and polarization DOFs in two-photon system assisted by a fixed frequency entangled state and a time intervals DOF.
The hyper-Bell states encoded in the spatial and polarization DOFs can be generated using the type-I two-crystal source spontaneous parametric down-conversion \cite{generation,six-state,frequency1,frequency2}.
The frequency entangled photon pairs can also be created via spontaneous parametric down-conversion or cold atoms  \cite{frequency00,cold-atoms,frequency01}, and it has been experimentally demonstrated \cite{frequency-experiment}.
The time intervals only need a longer optical circuit in experiment, which need no entanglement resource.
The FBS, which guides the photons  into the different
paths according to their frequencies, can be achieved by using  standard optical elements, a wavelength division multiplexer or fiber Bragg grating \cite{Standard-optical,Fiber Bragg Grating1,Fiber Bragg Grating2}. The FS, which is used to eliminate the frequency distinguishability,  can be realized by the frequency up-conversion process, or down-conversion process \cite{FS1,FS3,FS4}.
However, some unavoidable factors should also be considered in experiments, such as imperfect PBS, HWP, FBS, FS, and BS, imperfect hyperentanged state generation, dark count and background count of the detectors, drift in the interferometer during transmission, and phase miscalibration. In most HBSA schemes, one need to confirm the spatial mode of the photons before they move to the polarization state analysis. Our scheme is one-shot and there is no pause between each steps.

In Refs. \cite{assisted-DOF1,number-group1,Gao-incomplete}, the 16 hyper-Bell states are divided into 7, 12, and 14 discriminate groups, respectively. In our scheme, the 16 hyper-Bell states are distinguished unambiguously.
One fixed momentum (spatial) entangled state and one time-bin DOF are employed by Gao \emph{et al.} \cite{Gao-complete} to complete momentum-(spatial-) polarization HBSA.
One fixed momentum (spatial) entangled state and one fixed frequency entangled state are employed by Wang \emph{et al.} \cite{Wang-complete} to complete momentum- (spatial-) polarization HBSA.
One fixed time-bin entangled state and one fixed frequency entangled state  are employed by Zeng \emph{et al.}  \cite{Zeng-complete} to complete spatial-polarization HBSA.
Only one fixed frequency entangled state and time intervals DOF are employed in our scheme. Moreover, the frequency DOF suffers less from the channel noise in an optical fiber than polarization and momentum DOFs \cite{Fiber Bragg Grating1,Fiber Bragg Grating2,stable}.

In summary, we have proposed a scheme to completely distinguish 16 hyper-Bell states encoded in spatial and polarization DOFs assisted by the time intervals and a fixed frequency entanglement.
The parity information of the spatial and polarization DOFs can be distinguished resorting to the distinct time intervals of the photon-pair. The phase information of both the spatial and polarization DOFs in each group can be distinguished resorting to the detection signatures.
The scheme assisted by frequency and time interval DOFs, which is accessible to experiment with current technology. If the photon loss and the linear optics inefficiency are neglected, our scheme is deterministic.
Two auxiliary entangled states are necessary in Refs. \cite{Wang-complete,Zeng-complete}, while only one auxiliary entangled state is required in our scheme. Moreover, the frequency and time interval DOFs of photons suffer less from channel noise than spatial DOF. Our HBSA protocol maybe provide a potential tool to hyper-parallel one-way quantum computation, distributed quantum computation, and quantum communication.

\section*{Acknowledgments}\label{sec5}

This work is supported by the National Natural Science Foundation of China under Grant No. 11604012, the Fundamental Research Funds for the Central Universities under Grants FRF-TP-19-011A3, the Natural Science Foundation of China under Contract 61901420; the Shanxi Province Science Foundation for Youths under Contract 201901D211235; the Scientific and Technological Innovation Programs of Higher Education Institutions in Shanxi under Contract 2019L0507.

\vspace{6 cm}




\begin{thebibliography}{0}
\expandafter\ifx\csname natexlab\endcsname\relax\def\natexlab#1{#1}\fi
\expandafter\ifx\csname bibnamefont\endcsname\relax
  \def\bibnamefont#1{#1}\fi
\expandafter\ifx\csname bibfnamefont\endcsname\relax
  \def\bibfnamefont#1{#1}\fi
\expandafter\ifx\csname citenamefont\endcsname\relax
  \def\citenamefont#1{#1}\fi
\expandafter\ifx\csname url\endcsname\relax
  \def\url#1{\texttt{#1}}\fi
\expandafter\ifx\csname urlprefix\endcsname\relax\def\urlprefix{URL }\fi
\providecommand{\bibinfo}[2]{#2}
\providecommand{\eprint}[2][]{\url{#2}}

\end{thebibliography}


\begin{thebibliography}{99}


\bibitem{book} M. A. Nielsen and I. L. Chuang, \emph{Quantum Computation and Quantum Information} (Cambridge University, Cambridge, 2000).


\bibitem{one-way2} R. Raussendorf and H. J. Briegel, A one-way quantum computer, Phys. Rev. Lett. \textbf{86}, 5188 (2001).

\bibitem{one-way1} R. Raussendorf, D. E. Browne, and H. J. Briegel, Measurement-based quantum computation on cluster states, Phys. Rev. A \textbf{68}, 022312 (2003).







\bibitem{teleportation2} S. Liu, Y. Lou, and J. Jing, Orbital angular momentum multiplexed deterministic all-optical quantum teleportation, Nat. Commun. \textbf{11}, 3875 (2020).

\bibitem{teleportation3} S. Langenfeld, S. Welte, L. Hartung, S. Daiss, P. Thomas, O. Morin, E. Distante, and G. Rempe, Quantum teleportation between remote qubit memories with only a single photon as a resource, Phys. Rev. Lett. \textbf{126}, 130502 (2021).



\bibitem{swapping3} W. Ning, X. J. Huang, P. R. Han, H. Li, H. Deng, Z. B. Yang, Y. Xia, K. Xu, D. N. Zheng, and S. B. Zheng, Deterministic entanglement swapping in a superconducting circuit, Phys. Rev. Lett. \textbf{123}, 060502 (2019).


\bibitem{swapping2}Z. X. Ji, P. R. Fan, and H. G. Zhang, Entanglement swapping for Bell states and Greenberger-Horne-Zeilinger states in qubit systems, Physica A \textbf{585}, 126400 (2022).





\bibitem{QDC1}C. H. Bennett and S. J. Wiesner, Communication via one- and two-particle operators on Einstein-Podolsky-Rosen states, Phys. Rev. Lett. \textbf{69}, 2881 (1992).

\bibitem{QDC2}Y. Guo, B. H. Liu, C. F. Li, and G. C. Guo, Advances in quantum dense coding, Adv. Quantum Technol. \textbf{2}, 1900011 (2019).

\bibitem{QDC3} P. Wang, C. Q. Yu, Z. X. Wang, R. Y. Yuan, F. F. Du, and B. C. Ren, Hyperentanglement-assisted hyperdistillation for hyper-encoding photon system, Front. Phys.-Beijing \textbf{17}, 31501 (2022).





\bibitem{QKD6} G. L. Long and X. S. Liu, Theoretically efficient high-capacity quantum-key-distribution scheme, Phys. Rev. A \textbf{65},  032302 (2002).

\bibitem{QKD3} X. H. Li, F. G. Deng, and H. Y. Zhou, Efficient quantum key distribution over a collective noise channel, Phys. Rev. A \textbf{78}, 022321 (2008).

\bibitem{QKD5}L. M. Liang, S. H. Sun, M. S. Jiang, and C. Y. Li, Security analysis on some experimental quantum key distribution systems with imperfect optical and electrical devices, Front. Phys.-Beijing \textbf{9}, 613 (2014).

\bibitem{QKD4}C. C. W. Lim, F. Xu, J. W. Pan, and A. Ekert, Security analysis of quantum key distribution with small block length and its application to quantum space communications, Phys. Rev. Lett. \textbf{126}, 100501 (2021).

\bibitem{QKD-1} L. C. Kwek, L. Cao, W. Luo, Y. X. Wang, S. H. Sun, X. B. Wang, and A. Q. Liu, Chip-based quantum key distribution,  AAPPS Bull. \textbf{31}, 15 (2021).

\bibitem{QKD-2} C. Y. Gao, P. L. Guo, and B. C. Ren, Efficient quantum secure direct communication with complete Bell‐state measurement, Quant. Engineer. \textbf{3}, e83 (2021).

\bibitem{QKD-3} G. Z. Tang, C. Y. Li, and M. Wang, Polarization discriminated time-bin phase-encoding measurement-device-independent quantum key distribution, Quant. Engineer. \textbf{3}, e79 (2021).


\bibitem{EC1}X. Yan, Y. F. Yu, and Z. M. Zhang, Entanglement concentration for a non-maximally entangled four-photon cluster state, Front. Phys.-Beijing \textbf{9}, 640 (2014).

\bibitem{EC2}H. Wang, B. C. Ren, A. H. Wang, A. Alsaedi, T. Hayat, and F. G. Deng, General hyperentanglement concentration for polarization-spatial-time-bin multi-photon systems with linear optics, Front. Phys.-Beijing \textbf{13}, 130315 (2018).

\bibitem{EC3}J. Liu, L. Zhou, W. Zhong, and Y. B. Sheng, Logic Bell state concentration with parity check measurement, Front. Phys. \textbf{14}, 21601 (2019).



\bibitem{QEDC1} F. G. Deng, G. L. Long, and X. S. Liu, Two-step quantum direct communication protocol using the Einstein-Podolsky-Rosen pair block, Phys. Rev. A \textbf{68}, 042317 (2003).

\bibitem{QEDC2}Z. R. Zhou, Y. B. Sheng, P. H. Niu, L. G. Yin, G. L. Long, and L. Hanzo, Measurement-device-independent quantum secure direct communication, Sci. China Phys. Mech. Astron. \textbf{63}, 230362 (2020).

\bibitem{QEDC3} G. L. Long and H. Zhang, Drastic increase of channel capacity in quantum secure direct communication using masking, Sci. Bull. \textbf{66}, 1267 (2021).

\bibitem{QEDC4} Y. B. Sheng, L. Zhou, and G. L. Long, One-step quantum secure direct communication, Sci. Bull.  \textbf{67}, 367 (2022).

\bibitem{QEDC5} L. Zhou and Y. B. Sheng, One-step device-independent quantum secure direct communication, Sci. China Phys. Mech. Astron. \textbf{65}, 250311 (2022).

\bibitem{QEDC6} C. Wang, Quantum secure direct communication: Intersection of communication and cryptography, Funda. Res. \textbf{1}, 91 (2021).

\bibitem{QEDC7} Z. D. Ye, D. Pan, Z. Sun, C. G. Du, L. G. Yin, and G. L. Long, Generic security analysis framework for quantum secure direct communication, Front. Phys.-Beijing \textbf{16}, 21503 (2021).






\bibitem{impossible}N. L\"{u}tkenhaus, J. Calsamiglia, and K. A. Suominen, Bell measurements for teleportation, Phys. Rev. A \textbf{59}, 3295 (1999).

\bibitem{impossible1}L. Vaidman and N. Yoran, Methods for reliable teleportation, Phys. Rev. A \textbf{59}, 116 (1999).


\bibitem{NMR}M. A. Nielsen, E. Knill, and R. Laflamme, Complete quantum teleportation using nuclear magnetic resonance, Nature (London) \textbf{396}, 52 (1998).

\bibitem{atom} M. D. Barrett, J. Chiaverini, T. Schaetz, J. Britton, W. M. Itano, J. D. Jost, E. Knill, C. Langer, D. Leibfried, R. Ozeri, and D. J. Wineland, Deterministic quantum teleportation of atomic qubits, Nature (London) \textbf{429}, 737 (2004).

\bibitem{QED}L. Ye and G. C. Guo, Scheme for teleportation of an unknown atomic state without the Bell-state measurement, Phys. Rev. A \textbf{70}, 054303 (2004).



\bibitem{superdense} B. P. Williams, R. J. Sadlier, and T. S. Humble, Superdense coding over optical fiber links with complete Bell-state measurements, Phys. Rev. Lett. \textbf{118}, 050501 (2017).


\bibitem{assisted-DOF}P. G. Kwiat and H. Weinfurter, Embedded Bell-state analysis, Phys. Rev. A \textbf{58}, R2623 (1998).


\bibitem{hyperen1} G. Vallone, R. Ceccarelli, F. De Martini, and P. Mataloni, Hyperentanglement of two photons in three degrees of freedom, Phys. Rev. A \textbf{79}, 030301 (2009).

\bibitem{hyperen2} X. L. Wang, Y. H. Luo, H. L. Huang, M. C. Chen, Z. E. Su, C. Liu, C. Chen, W. Li, Y. Q. Fang, X. Jiang, J. Zhang, L. Li, N. L. Liu, C. Y. Lu, and J. W. Pan, 18-qubit entanglement with six photons' three degrees of freedom, Phys. Rev. Lett. \textbf{120}, 260502 (2018).




\bibitem{photon-DOF2}D. Pile, How many bits can a photon carry? Nat. Photon. \textbf{6}, 14 (2012).

\bibitem{photon-DOF1} F. Flamini, N. Spagnolo, and F.  Sciarrino, Photonic quantum information processing: A review, Rep. Prog. Phys. \textbf{82}, 016001 (2018).



\bibitem{nonlinear8}G. Y. Wang, Q. Ai, B. C. Ren, T. Li, and F. G. Deng, Error-detected generation and complete analysis of hyperentangled Bell states for photons assisted by quantum-dot spins in double-sided optical microcavities, Opt. Express \textbf{24}, 28444 (2016).

\bibitem{application} F. G. Deng, B. C. Ren, and X. H. Li, Quantum hyperentanglement and its applications in quantum information processing, Sci. Bull. \textbf{62}, 46 (2017).


\bibitem{assisted-DOF1}T. C. Wei,  J. T. Barreiro, and P. G. Kwiat, Hyperentangled Bell-state analysis, Phys. Rev. A \textbf{75}, 060305(R) (2007).

\bibitem{seven1}N. Pisenti,  C. P. E. Gaebler, and T. W. Lynn, Distinguishability of hyperentangled Bell states by linear evolution and local projective measurement, Phys. Rev. A \textbf{84}, 022340 (2011).




\bibitem{number-group1}X. H. Li and S. Ghose, Hyperentangled Bell-state analysis and hyperdense coding assisted by auxiliary entanglement, Phys. Rev. A \textbf{96}, 020303(R) (2017).

\bibitem{Gao-complete}C. Y. Gao, B. C. Ren, Y. X. Zhang, Q. Ai, and F. G. Deng, The linear optical unambiguous discrimination of hyperentangled Bell states assisted by time bin, Ann. Phys. (Berlin) \textbf{531}, 1900201 (2019).





\bibitem{Gao-incomplete} C. Y. Gao, B. C. Ren, Y. X. Zhang, Q. Ai, and F. G. Deng, Universal linear-optical hyperentangled Bell-state measurement, Appl. Phys. Express \textbf{13}, 027004 (2020).




\bibitem{Kerr} Y. B. Sheng, F. G. Deng, and G. L. Long, Complete hyperentangled-Bell-state analysis for quantum communication, Phys. Rev. A \textbf{82}, 032318 (2010).

\bibitem {nonlinear6}Y. B. Sheng and L. Zhou, Two-step complete polarization logic Bell-state analysis, Sci. Rep. \textbf{5}, 13453 (2015).


\bibitem{nonlinear7}X. H. Li and S. Ghose, Self-assisted complete maximally hyperentangled state analysis via the cross-Kerr nonlinearity, Phys. Rev. A \textbf{93}, 022302 (2016).

\bibitem{Zhang-kerr} H. R. Zhang, P. Wang, C. Q. Yu, and B. C. Ren, Deterministic nondestructive state analysis for polarization-spatial-time-bin hyperentanglement with cross-Kerr nonlinearity, Chin. Phys. B \textbf{30}, 030304 (2021). 


\bibitem {nonlinear3} B. C. Ren, H. R. Wei, M. Hua, T. Li, and F. G. Deng, Complete hyperentangled-Bell-state analysis for photon systems assisted by quantum-dot spins in optical microcavities, Opt. Express \textbf{20}, 24664 (2012).

\bibitem{nonlinear4} T. J. Wang, Y. Lu, and G. L. Long, Generation and complete analysis of the hyperentangled Bell state for photons assisted by quantum-dot spins in optical microcavities, Phys. Rev. A \textbf{86}, 042337 (2012).

\bibitem{nonlinear5}Q. Liu and M. Zhang, Generation and complete nondestructive analysis of hyperentanglement assisted by nitrogen-vacancy centers in resonators, Phys. Rev. A \textbf{91}, 062321 (2015).


\bibitem{superconductor}E. Sabag, S. Bouscher, R. Marjieh, and A. Hayat, Photonic Bell-state analysis based on semiconductor-superconductor structures, Phys. Rev. B \textbf{95}, 094503 (2017).

\bibitem{nonlinear11}Y. Y. Zheng, L. X. Liang, and M. Zhang, Error-heralded generation and self-assisted complete analysis of two-photon hyperentangled Bell states through single-sided quantum-dot-cavity systems, Sci. China Phys. Mech. \& Astron. \textbf{62}, 970312 (2019).

\bibitem{nonlinear12}C. Cao, L. Zhang, Y. H. Han, P. P. Yin, L. Fan, Y. W. Duan, and R. Zhang, Complete and faithful hyperentangled-Bell-state analysis of photon systems using a failure-heralded and fidelity-robust quantum gate, Opt. Express  \textbf{28}, 2857 (2020). 


\bibitem{Wang-complete} G. Y. Wang, B. C. Ren, F. G. Deng, and G. L. Long, Complete analysis of hyperentangled Bell states assisted with auxiliary hyperentanglement, Opt. Express \textbf{27}, 8994 (2019).

\bibitem{Zeng-complete} Z. Zeng and K. D. Zhu, Complete hyperentangled Bell state analysis assisted by hyperentanglement, Laser Phys. Lett. \textbf{17}, 075203 (2020).





\bibitem{linear time}T. Li, G. Y. Wang, F. G. Deng, and G. L. Long, Deterministic error correction for nonlocal spatial-polarization hyperentanglement, Sci. Rep. \textbf{6}, 20677 (2016).


\bibitem{generation}P. G. Kwiat, E. Waks, A. G. White, I. Appelbaum, and P. H. Eberhard, Ultrabright source of polarization-entangled photons, Phys. Rev. A \textbf{60}, R773 (1999).

\bibitem{six-state} R. Ceccarelli, G. Vallone, F. De Martini, P. Mataloni, and A. Cabello,  Experimental entanglement and nonlocality of a two-photon six-qubit cluster state, Phys. Rev. Lett. \textbf{103}, 160401 (2009).



\bibitem{frequency1}A. Yabushita and T. Kobayashi, Spectroscopy by frequency-entangled photon pairs, Phys. Rev. A \textbf{69}, 013806 (2004).

\bibitem{frequency2}A. Yabushita and T. Kobayashi, Generation of frequency tunable polarization entangled photon pairs, J. Appl. Phys. \textbf{99}, 063101 (2006).


\bibitem{cold-atoms} C. Shu, X. X. Guo, P. Chen, M. M. T. Loy, and S. W. Du, Narrowband biphotons with polarization-frequency-coupled entanglement, Phys. Rev. A \textbf{91}, 043820 (2015).


\bibitem{frequency00} W. Ueno, F. Kaneda, H. Suzuki, S. Nagano, A. Syouji, R. Shimizu, K. Suizu, and K. Edamatsu,
 Entangled photon generation in two-period quasi-phase-matched parametric down-conversion, Opt. Express \textbf{20}, 5508 (2012).


\bibitem{frequency01} F. Kaneda, H. Suzuki, R. Shimizu, and K. Edamatsu, Direct generation of frequency-bin entangled photons via two-period quasi-phase-matched parametric downconversion, Opt. Express \textbf{27}, 1416 (2019).




\bibitem{frequency-experiment} A. V. Burlakov, S. P. Kulik, G. O. Rytikov, and  M. V. Chekhova, Biphoton light generation in polarization-frequency Bell states,
J. Exp. Thror. Phys. \textbf{95}, 639 (2002).
%



\bibitem{Standard-optical} E. H. Huntington and T. C. Ralph, Components for optical qubits encoded in sideband modes, Phys. Rev. A \textbf{69}, 042318 (2004).


\bibitem{Fiber Bragg Grating1} M. Bloch, S. W. McLaughlin, J. M. Merolla, and F. Patois, Frequency-coded quantum key distribution, Opt. Lett. \textbf{32}, 301 (2007).

\bibitem{Fiber Bragg Grating2}T. Zhang, Z. Q. Yin, Z. F. Han, and G. C. Guo, A frequency-coded quantum key distribution scheme, Opt. Commun. \textbf{281}, 4800 (2008).


\bibitem{FS1} C. Langrock, E. Diamanti, R. V. Roussev, Y. Yamamoto, M. M. Fejer, and H. Takesue, Highly efficient single-photon detection at communication wavelengths by use of upconversion in reverse-proton-exchanged periodically poled LiNbO$_3$ waveguides, Opt. Lett. \textbf{30}, 1725 (2005). 




\bibitem{FS3} H. Takesue, Erasing distinguishability using quantum frequency up-conversion, Phys. Rev. Lett. \textbf{101}, 173901 (2008). 


\bibitem{FS4} R. Ikuta, Y. Kusaka, T. Kitano, H. Kato, T. Yamamoto, M. Koashi, and N. Imoto, Wide-band quantum interface for visible-to-telecommunication wavelength conversion, Nat. Commun. \textbf{2}, 537 (2011). 





\bibitem{stable} J. M. Merolla, L. Duraffourg, J. P. Goedgebuer, A. Soujaeff, F. Patois and W. T. Rhodes, Integrated quantum key distribution system using single sideband detection, Eur. Phys. J. D \textbf{18}, 141 (2002).



\end{thebibliography}
\end{document}